\documentclass[12pt,preprint]{aastex}

\shorttitle{\indent \def Cool and hot components of a coronal
bright point} \shortauthors{Tian et al.}

\begin{document}

\title{Cool and hot components of a coronal bright point}

\author{Hui Tian\altaffilmark{1,2}, Werner Curdt\altaffilmark{1}, Eckart Marsch\altaffilmark{1}, Jiansen He\altaffilmark{2}}
\altaffiltext{1}{Max-Planck-Institut f\"ur Sonnensystemforschung,
37191 Katlenburg-Lindau, Germany; tianhui924@gmail.com}

\altaffiltext{2}{Department of Geophysics, Peking University,
100871, Beijing, China}

\begin{abstract}
We performed a systematic study of the Doppler shifts and electron
densities measured in an EUV bright point (hereafter BP) observed in
more than 10 EUV lines with formation temperatures from
$\log(T/\rm{K}) = 4.5$ to 6.3. Those parts of a BP seen in
transition region and coronal lines are defined as its cool and hot
components, respectively. We find that the transition from cool to
hot occurs at a temperature around $\log(T/\rm{K}) = 5.7$. The two
components of the BP reveal a totally different orientation and
Doppler-shift pattern, which might result from a twist of the
associated magnetic loop system. The analysis of magnetic-field
evolution and topology seems to favour a two-stage heating process,
in which magnetic cancelation and separator reconnection are
powering, respectively, the cool and hot components of the BP. We
also found that the electron densities of both components of the BP
are higher than those of the surrounding quiet Sun, and comparable
to or smaller than active-region densities.
\end{abstract}

\keywords{Sun: solar corona, transition region, UV radiation}

\section{Introduction}
Coronal bright points (BPs) are small-scale phenomena in the solar
corona and characterized by enhanced emission in X-ray,
extreme-ultraviolet (EUV) and radio wavelengths. They were found to
be located at the network boundaries, where the quiet-Sun magnetic
field is mainly concentrated \citep{HabbalEtal1990} and to expand
with height \citep{Gabriel1976,TianEtal2008}. Coronal BPs are
typically $30^{\prime\prime}-40^{\prime\prime}$ in size, often with
a bright core of $5^{\prime\prime}-10^{\prime\prime}$
\citep{MadjarskaEtal2003}. It is believed that a BP consists of
several miniature dynamic loops \citep{SheeleyGolub1979}. The
average lifetime of BPs is 20 hours in EUV \citep{ZhangEtal2001} and
8 hours in X-ray observations \citep{GolubEtal1974}.

\cite{MadjarskaEtal2003} found that the Doppler shift of S~{\sc{vi}}
in a BP is in the range of $-10$ to 10~km/s. \cite{XiaEtal2003}
found that the BPs in a coronal hole corresponded to small blue
shift of Ne~{\sc{viii}}. Recently, both up and down flows were
detected in a BP \citep{BrosiusEtal2007}. The electron densities of
BPs have been derived by using line pairs with a formation
temperature of $\log(T/\rm{K})=6.1-6.3$
\citep{UgarteEtal2005,BrosiusEtal2008}.

The evolution of a BP is strongly related with the underlying
bipolar magnetic field
\citep{WebbEtal1993,FalconerEtal1998,BrownEtal2001,MadjarskaEtal2003,TianEtal2007}.
Most BPs are more likely associated with the cancellation of
magnetic features than their emergence \citep{WebbEtal1993}. The
energization of BPs may result from the interaction between two
magnetic fragments of opposite polarities
\citep{PriestEtal1994,ParnellEtal1994,VonEtal2006}, magnetic
reconnection along separator field lines \citep{Longcope1998} or
current sheets induced by photospheric
motions\citep{Buchner2006,Santos2007}.

Here we study the Doppler shifts and electron densities of a BP,
which was observed by spectrometers onboard SOHO and \emph{Hinode}
in a wide temperature range. We will show the differences in the
morphology and Doppler pattern between the cool and hot components
of this BP, and discuss possible mechanisms powering the radiation
of the two components.

\section{Data analysis}

The data set analysed here was obtained by the Solar Ultraviolet
Measurements of Emitted Radiation instrument
(SUMER)~\citep{WilhelmEtal1995,LemaireEtal1997} and the EUV Imaging
Spectrometer (EIS)~\citep{CulhaneEtal2007} on 5 April 2007. From
01:44 to 05:13 UTC, SUMER three times scanned a quiet Sun region
around disk center with a size of about
$58^{\prime\prime}\times120^{\prime\prime}$, with each scan using
different spectral windows. Slit 4
($1^{\prime\prime}\times120^{\prime\prime}$) was used and the
exposure time was 90~s. The raster increment was about
$1.5^{\prime\prime}$. EIS scanned from 03:15 to 04:09 UTC almost the
same region by using the $1^{\prime\prime}$ slit with an exposure
time of 90~s. The raster increment was about $2^{\prime\prime}$. The
coalignment between different SUMER images were done by using the
cross-correlation technique. The He~{\sc{ii}}~(256.32{\AA}) image
taken by EIS was used to coalign SUMER and EIS. During this period,
an EUV BP was present in the field of view of the spectrometers. We
selected some strong and clean spectral lines (see
Table~\ref{table1}) to study the Doppler shifts, and some
density-sensitive line pairs (see Fig.~\ref{fig.4}) to study the
electron density of this BP. The values of rest wavelengths and
formation temperatures were taken from \cite{Xia2003} and the
CHIANTI data base \citep{DereEtal1997,LandiEtal2006}.

We applied the standard procedures for correcting and calibrating
the SUMER data, including decompression, flat-field correction, and
detector corrections for geometrical distortion, local gain and dead
time. We also used the standard EIS calibrating procedures to
subtract the dark-current, remove the effects of cosmic rays and hot
pixels, and to make wavelength and absolute calibrations.

In order to build up intensity maps and Dopplergrams for the lines
listed in Table~\ref{table1}, we applied a single Gaussian fit to
each spectral profile. Then we estimated the line shift caused by
the thermal deformation of the optical system of SUMER, using a
similar method as described in \cite{DammaschEtal1999}. Also, the
fitted center of an EIS line was further corrected by taking into
account the slit-tilt effect and orbital variation of the line
position. We assumed a vanishing Doppler shift for each line when
being averaged over the whole FOV. In this way we could calibrate
the wavelength and obtained a Dopplergram for each line.
Fig.~\ref{fig.1} and Fig.~\ref{fig.2} show part of the resulting
intensity maps and Dopplergrams, respectively.

During this time period, only the 96-min MDI (Michelson Doppler
Imager) magnetograms \citep{ScherrerEtal1995} were available.
Fig.~\ref{fig.3} shows four magnetograms of the region in the
vicinity of the BP taken at four different times. They were
coaligned by using the SUMER continuum intensity image obtained
around 750~{\AA}. The right-most panel of Fig.~\ref{fig.3} is the
magnetic skeleton of the BP region projected onto the plane of the
photosphere. We first used the YAFTA (Yet Another Feature Tracking
Algorithm) software \citep{WelschEtal2004,DeForestEtal2007} to
select all individual magnetic concentrations and then reduced them
to point sources. With the help of the MPOLE (the Magnetic Charge
Topology and Minimum Current Corona model) software
\citep{LongcopeKlapper2002}, we could reconstruct the potential
magnetic field topology produced by the set of point sources.

Some lines listed in Fig.~\ref{fig.4} are weak or blended. However,
by using the method of single or double gaussian fitting, we can
still obtain a reliable intensity for each line in the BP region.
Each of the density-sensitive line pairs was observed
simultaneously. The theoretical relations between intensity ratios
of line pairs and electron densities are shown in the top of
Fig.~\ref{fig.4} and were taken from the CHIANTI data base
\citep{DereEtal1997,LandiEtal2006}. Some blends discussed in
\cite{YoungEtal2007} were also considered. We calculated intensities
and densities for each individual pixel within the corresponding
contoured region in Fig.~\ref{fig.2} and then averaged those
densities. We should mention that the O~{\sc{iv}} line pair was
observed at about 11:40, when the BP was still there. The results
are shown in the bottom of Fig.~\ref{fig.4}.

\section{Results and discussion}

The temperature dependence of the emission of BPs and their
different plasma properties at different wavelengths are not yet
well understood. Most of the previous studies concentrated on BPs as
seen in only one wavelength or band. Few authors have analyzed a BP
seen simultaneously at different temperatures
\citep{HabbalEtal1981,HabbalEtal1990,HabbalEtal1991,UgarteEtal2004,McIntosh2007,BrosiusEtal2007}.

\begin{figure}
\centering {\includegraphics[width=0.9\textwidth]{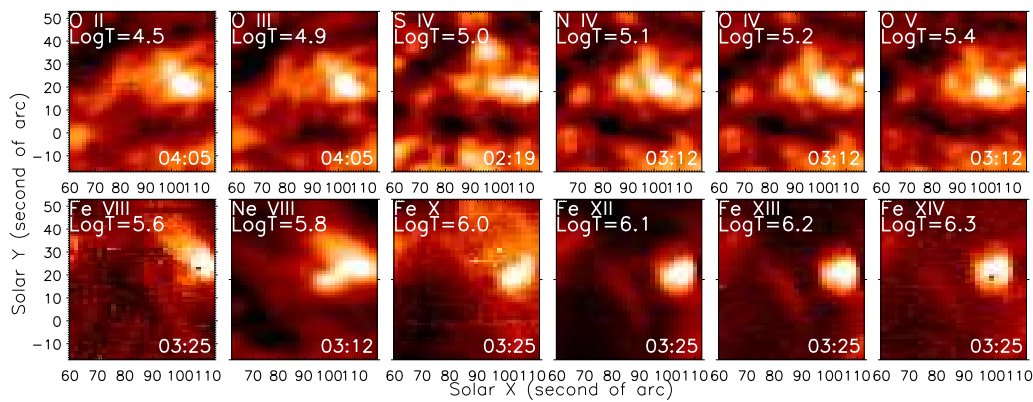}}
\caption{Intensity maps (on a logarithmic scale) of the lines listed
in Table~\ref{table1}. The approximate time when the BP was scanned
is shown in the lower right corner of each map.} \label{fig.1}
\end{figure}

\begin{figure}
\centering {\includegraphics[width=0.9\textwidth]{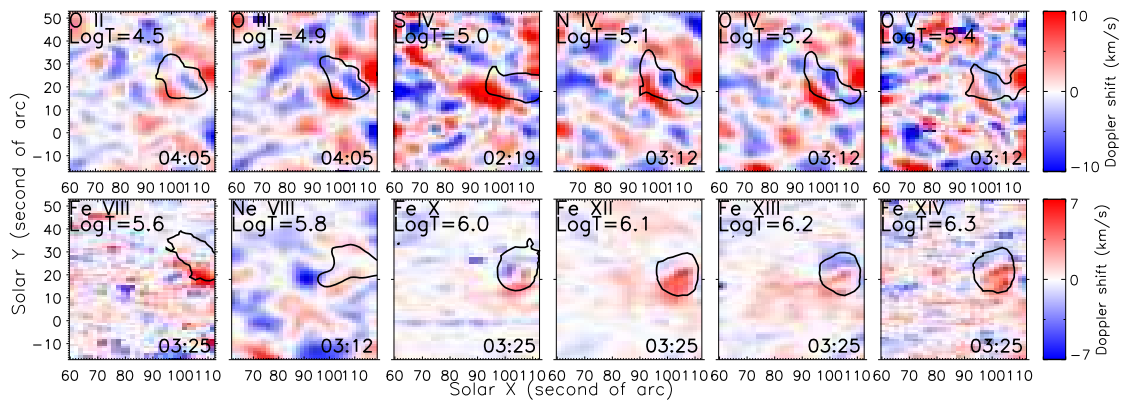}}
\caption{Dopplergrams of the lines listed in Table~\ref{table1}. The
approximate time when the BP was scanned is shown in the lower right
corner of each map. The black contours outline the positions of the
BP as seen in different wavelengths.} \label{fig.2}
\end{figure}

Our Fig.~\ref{fig.1} reveals the rich morphology of a BP when
observed with such a wide temperature coverage. The Dopplergrams and
related intensity contours are shown in Fig.~\ref{fig.2}. It is
believed that BPs can experience variations in emission intensity
and Doppler shift on a time scale of minutes
\citep{HabbalEtal1990,MadjarskaEtal2003}. However, the general
emission pattern of some BPs can remain almost the same for 1-3
hours (see Fig.~\ref{fig.2} in \cite{BrownEtal2001}). We should not
exclude the possibility of the presence of a more or less stable
component of emission and Doppler flow in some BPs. In our case, the
orientation and Doppler pattern (especially the sharp boundary
between patches of red and blue shift) of the BP are both different
at higher ($\log(T/\rm{K})>5.7$) and lower ($\log(T/\rm{K})<5.7$)
temperatures. At lower temperatures, although the BP was observed at
different times from 02:19 to 04:05, its orientation and Doppler
pattern seem to be consistent and do not change too much.

Recently, \cite{BrosiusEtal2007} found Doppler shifts on opposite
sides of a BP ranging from $-15$ to $+15$~km/s in He~{\sc{ii}}, and
from $-35$ to $+35$~km/s in Fe~{\sc{xvi}}. Our results demonstrate
that the Doppler shifts in the BP are quite different in different
lines. Patches of red and blue shift are both found in the BP with
comparable sizes. The absolute shift is largest in middle-transition
region lines ($\log(T/\rm{K})=4.9-5.2$) and can reach more than
10~km/s.

The most interesting phenomenon visible in Fig.~\ref{fig.2} is that
the boundary between up/down flows in the BP is totally different,
i.e. almost perpendicular, for lines with lower and higher
temperatures. This boundary in Ne~{\sc{viii}} resembles those of
lines with a higher temperature. But the Doppler pattern of
Ne~{\sc{viii}} is similar to those of lines with a lower temperature
in the surrounding quiet Sun. \cite{McIntosh2007} classified the BPs
seen in He~{\sc{ii}}~(304~{\AA}) and soft X-ray as cool and hot BPs,
respectively. Similarly, we define here the bright feature seen at
coronal temperature as the hot component, and the corresponding
bright emission at TR temperature as the cool component of a BP.
Considering the morphology and Doppler pattern of the BP, we found
that a transition from the cool to the hot component of the BP
occurs at a temperature of about $\log(T/\rm{K})=5.7$.

\begin{figure}
\centering
{\includegraphics[height=0.13\textheight,width=0.9\textwidth]{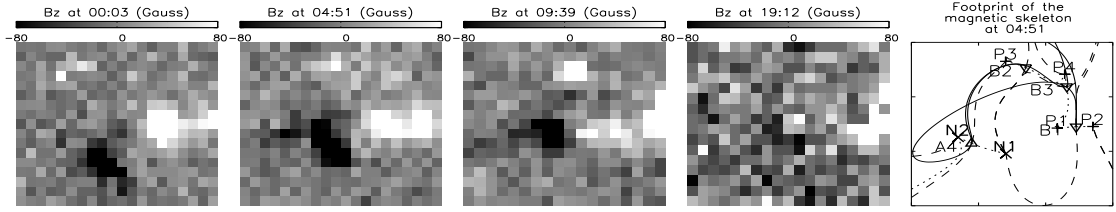}}
\caption{From left to right: MDI magnetograms taken at 00:03, 04:51,
09:39, 19:12, and the magnetic skeleton at 04:51. The FOV
corresponds to a coordinate range of x from $80^{\prime\prime}$ to
$115^{\prime\prime}$, and of y from $5^{\prime\prime}$ to
$37^{\prime\prime}$. In the magnetic skeleton, positive/negative
magnetic point sources are labelled P/N and marked with plus/cross
symbols; positive/negative nulls are labelled B/A and marked with
triangles pointing down/up; spines and separators are shown as
dotted and solid lines,respectively; dashed lines indicate where
separatrix surfaces intersect the photosphere.} \label{fig.3}
\end{figure}

The different boundary between up/down flows at lower and higher
temperatures in the BP might be a result of a syphon flow along a
twisted loop system, which twists or spirals at its upper segment.
Then the flow may lead to comparable red and blue shifts in the
opposite legs, and result in a different boundary of the up and down
flows located between the lower and upper parts of the loop. In
Fig.~\ref{fig.3}, we found that the main negative magnetic source
rotated relative to the positive source in a right-hand sense during
the time period from 00:03 to 09:39. This movement might indicate a
continual twist of the entire loop system. We have applied the
force-free magnetic field extrapolation technique
\citep{Seehafer1978} to reconstruct the 3-D magnetic field in our BP
region, but we could not reconstruct such a helical loop system.
This does not mean that the twisted loop is not there, because the
BP may simply not be a force-free structure, and consequently the
extrapolation used may not represent the real magnetic field
adequately. However, the up and down flows in our case might also be
associated with magnetic reconnection as discussed in
\cite{BrosiusEtal2007}.

The different emission morphology and Doppler pattern between the
cool and hot components of the BP may imply a different powering
mechanism of the two components. \cite{McIntosh2007} proposed a
two-stage heating process, in which magnetoconvection-driven
reconnection occurs in and supplies energy to the cool BPs,
whereupon the increased energy supply leads to an expansion of the
loop system, which interacts with the overlying coronal magnetic
flux through fast separator reconnection and produces hot BPs. From
magnetograms taken before and after the observation periods of SUMER
and EIS, we find that the main negative source in the left frames of
Fig.~\ref{fig.3} is seen to approach and partly cancel the positive
source. This cancellation might have started before, and still be in
process after our spectroscopic observations. We also find that the
three separators (A4-B1,A4-B2,A4-B3) in the magnetic skeleton at
04:51 partly fit the orientation of the hot component of the BP.
Thus, our observations seem to support the two-stage powering
mechanism.

\begin{figure}
\centering
{\includegraphics[height=0.4\textheight,width=7.0cm]{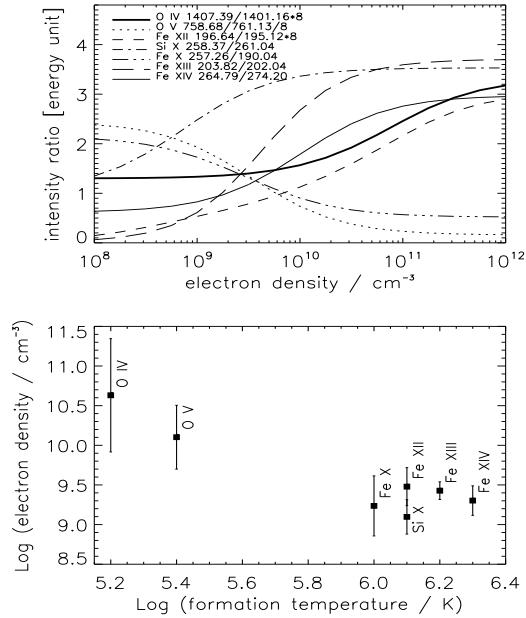}}
\caption{Top: The theoretical relation between the intensity ratios
of 6 line pairs and the corresponding electron densities. Bottom:
The derived electron densities and their uncertainties indicated by
standard deviation bars.} \label{fig.4}
\end{figure}

The electron density is vital to determine the radiative losses in
coronal heating models. Here we derived separately the densities of
the cool and hot components of the same BP. The results are
summarized in Fig.~\ref{fig.4}. Our derived values of the electron
density in the range from $\log(T/\rm{K})=6.1$ to $6.3$ are similar
to those of \cite{UgarteEtal2005} and \cite{BrosiusEtal2008}. These
densities are in the range of values typical of ARs, and slightly
larger than quiet-Sun densities. The result of the higher value
obtained with Fe~{\sc{xii}} compared to Si~{\sc{x}} is nothing new
and a related discussion can be found in \cite{UgarteEtal2005}.

The densities of the cool component of the BP have large
uncertainties. But they can still be considered to be much higher
than the density of the normal quiet Sun, which has an upper limit
of $\log(N_e/\rm{cm}^{-3})=9.87$ at $\log(T/\rm{K})=5.25$ in
\cite{GriffithsEtal1999}. \cite{TripathiEtal2008} derived a density
of $\log(N_e/\rm{cm}^{-3})=10.0-10.5$ at $\log(T/\rm{K})=5.8-6.1$
inside the moss region of an AR. If we assume the density decreases
with increasing temperature in the transition region, we can expect
an even higher density of the AR at $\log(T/\rm{K})=5.2-5.4$. It
should be comparable to, or larger than, our derived densities at
$\log(T/\rm{K})=5.2-5.4$. So, the conclusion in
\cite{UgarteEtal2005}, namely that the BP plasma has more
similarities to active-region than quiet-Sun plasma, may also be
true for the cool component.

\acknowledgements SUMER and MDI are instruments onboard SOHO, an ESA
and NASA mission. The SUMER project is financially supported by DLR,
CNES, NASA, and the ESA PRODEX programme (Swiss contribution). EIS
is an instrument onboard {\it Hinode}, a Japanese mission developed
and launched by ISAS/JAXA, with NAOJ as domestic partner and NASA
and STFC (UK) as international partners. It is operated by these
agencies in co-operation with ESA and NSC (Norway). The work of Hui
Tian and Jiansen He's team in PKU is supported by the National
Natural Science Foundation of China under contracts 40574078 and
40436015. Hui Tian is now supported by China Scholarship Council for
his stay in Germany. We thank the anonymous referee and Dr. M.
Madjarska for the helpful comments.

\clearpage

\begin{table}[]
\caption[]{ Emission lines used to study the Doppler shifts of the BP. Here
$\lambda$ and $T$ represent the rest wavelength and formation temperature,
respectively.}
\label{table1}
\begin{center}
\begin{tabular}{p{1.0cm} p{1.2cm} p{1.5cm}| p{1.0cm} p{1.1cm} p{1.5cm}}
\hline\hline \multicolumn{3}{c}{SUMER lines} \vline
&\multicolumn{3}{c}{EIS lines}\\
\hline Ion & $\lambda$ ({\AA}) &  $\log(T/\rm{K})$
     & Ion & $\lambda$ ({\AA}) &  $\log(T/\rm{K})$ \\
\hline
O~{\sc{ii}} & 833.332 &   4.5 &Fe~{\sc{viii}}& 185.21  &  5.6 \\
O~{\sc{iii}}& 833.749 &   4.9 &Fe~{\sc{x}}   & 184.54  &  6.0 \\
S~{\sc{iv}} & 750.221 &   5.0 &Fe~{\sc{xii}} & 195.12  &  6.1 \\
N~{\sc{iv}} & 765.148 &   5.1 &Fe~{\sc{xiii}}& 202.04  &  6.2 \\
O~{\sc{iv}} & 787.711 &   5.2 &Fe~{\sc{xiv}} & 264.79  &  6.3 \\
O~{\sc{v}}  & 758.677 &   5.4 &    &    & \\
Ne~{\sc{viii}}& 770.428 & 5.8 &    &    & \\
\hline
\end{tabular}
\end{center}
\end{table}

\clearpage

\begin{thebibliography}{}
\bibitem[Brosius et al.(2007)]{BrosiusEtal2007}
Brosius, J. W., Rabin, D. M., \& Thomas, R. J. 2007, ApJ, 656, L41
%
\bibitem[Brosius et al.(2008)]{BrosiusEtal2008}
Brosius, J. W., Rabin, D. M., Thomas, R. J., \& Landi, E. 2008, ApJ,
677, 781
%
\bibitem[Brown et al.(2001)]{BrownEtal2001}
Brown, D. S., et al. 2001, Sol. Phys., 201, 305
%
\bibitem[B\"{u}chner (2006)]{Buchner2006}
B\"{u}chner, J. 2006, Space Science Reviews, 122, 149
%
\bibitem[Culhane et al.(2007)]{CulhaneEtal2007}
Culhane, J.L. et al. 2007, Solar Phys., 243, 19
%
\bibitem[Dammasch et al.(1999)]{DammaschEtal1999}
Dammasch, I. E., Wilhelm, K., Curdt, W.,\& Hassler, D. M. 1999, A\&A, 346, 285
%
\bibitem[DeForest et al.(2007)]{DeForestEtal2007}
DeForest, C. E., et al. 2007, ApJ, 666, 576
%
\bibitem[Dere et al.(1997)]{DereEtal1997}
Dere, K. P., et al. 1997, A\&AS, 125, 149
%
\bibitem[Gabriel(1976)]{Gabriel1976}
Gabriel A. H. 1976, Philos. Trans. R. Soc. London A, 281, 575
%
\bibitem[Falconer et al.(1998)]{FalconerEtal1998}
Falconer, D. A., Moore, R. L., Porter, J. G., \& Hathaway, D. H. 1998, ApJ, 501, 386
%
\bibitem[Golub et al.(1974)]{GolubEtal1974}
Golub, L., et al. 1974, ApJ, 189, L93
%
\bibitem[Griffiths et al.(1999)]{GriffithsEtal1999}
Griffiths, N. W., Fisher, G. H., Woods, D. T., \& Siegmund,O. H. W. 1999, ApJ, 512,992
%
\bibitem[Habbal and Withbroe(1981)]{HabbalEtal1981}
Habbal, S. R., \& Withbroe, G. L. 1981, Sol. Phys., 69, 77
%
\bibitem[Habbal et al.(1990)]{HabbalEtal1990}
Habbal. S. R., Dowdy, J. F. Jr., \&Withbroe, G. L. 1990, ApJ, 352,333
%
\bibitem[Habbal and Grace(1991)]{HabbalEtal1991}
Habbal, S. R., \& Grace, E. 1991, ApJ, 382, 667
%
\bibitem[Landi et al.(2006)]{LandiEtal2006}
Landi, E., et al. 2006, ApJS, 162, 261
%
\bibitem[Lemaire et al.(1997)]{LemaireEtal1997}
Lemaire P., Wilhelm K., Curdt W., et al. 1997, Sol. Phys., 170,105
%
\bibitem[Longcope(1998)]{Longcope1998}
Longcope, D. W. 1998, ApJ, 507, 433
%
\bibitem[Longcope and Klapper(2002)]{LongcopeKlapper2002}
Longcope, D. W. \& Klapper, I. 2002, ApJ, 579, 468
%
\bibitem[Madjarska et al.(2003)]{MadjarskaEtal2003}
Madjarska, M. S., Doyle, J. G., Teriaca, L., \& Banerjee, D. 2003, A\&A, 398, 775
%
\bibitem[McIntosh(2007)]{McIntosh2007}
McIntosh, S. W. 2007, ApJ, 670, 1401
%
\bibitem[Parnell et al.(1994)]{ParnellEtal1994}
Parnell, C. E., Priest, Eric R., \&Titov, V. S. 1994, Sol. Phys., 153, 217
%
\bibitem[Priest et al.(1994)]{PriestEtal1994}
Priest, E. R., Parnell, C. E., \&Martin, S. F. 1994, ApJ, 427, 459
%
\bibitem[Santos and B\"{u}chner(2007)]{Santos2007}
Santos, J. C., \&B\"{u}chner, J., Astrophys. Space Sci. Trans. 2007,
3, 29
%
\bibitem[Scherrer et al.(1995)]{ScherrerEtal1995}
Scherrer, P. H., et al. 1995, Solar Phys., 162,129
%
\bibitem[Seehafer(1978)]{Seehafer1978}
Seehafer, N. 1978, Sol. Phys., 58, 215
%
\bibitem[Sheeley and Golub(1979)]{SheeleyGolub1979}
Sheeley, N. R. Jr., \& Golub, L. 1979, Sol. Phys., 63, 119
%
\bibitem[Tian et al.(2007)]{TianEtal2007}
Tian H., Tu C.-Y., He J.-S., \&Marsch, E. 2007, Adv. Space Res., 39,1853
%
\bibitem[Tian et al.(2008)]{TianEtal2008}
Tian H., Marsch E., Tu C.-Y., Xia L.-D., \&He J.-S. 2008, A\&A, 482,267
%
\bibitem[Tripathi et al.(2008)]{TripathiEtal2008}
Tripathi, D., Mason, H. E., Young, P. R., \& Del Zanna, G. 2008, A\&A 481, L53
%
\bibitem[Ugarte-Urra et al.(2004)]{UgarteEtal2004}
Ugarte-Urra, I., Doyle, J. G., Madjarska, M. S., \& O$^{\prime}$Shea, E. 2004, A\&A, 418, 313
%
\bibitem[Ugarte-Urra et al.(2005)]{UgarteEtal2005}
Ugarte-Urra, I., Doyle, J. G., \& Del Zanna G. 2005, A\&A, 435, 1169
%
\bibitem[Von Rekowski et al.(2006)]{VonEtal2006}
Von Rekowski B., Parnell, C. E., \&Priest, E. R. 2006, MNRAS, 366, 125
%
\bibitem[Webb et al.(1993)]{WebbEtal1993}
Webb, D. F., Martin, S. F., Moses, D., \& Harvey, J. W. 1993, Sol. Phys., 144, 15
%
\bibitem[Welsch et al.(2004)]{WelschEtal2004}
Welsch, B. T., Fisher, G. H., Abbett, W. P., \& Regnier, S. 2004, ApJ, 610, 1148
%
\bibitem[Wilhelm et al.(1995)]{WilhelmEtal1995}
Wilhelm K., Curdt W., Marsch E., et al. 1995, Sol. Phys., 162,189
%
\bibitem[Xia et al.(2003)]{XiaEtal2003}
Xia, L.~D., Marsch, E., \& Curdt, W. 2003, A \& A, 399, L5
%
\bibitem[Xia(2003)]{Xia2003}
Xia, L.-D. 2003, Ph.D. Thesis (G\"ottingen: Georg-August-Univ.)
%
\bibitem[Young et al.(2007)]{YoungEtal2007}
Young,P. R., et al. 2007, PASJ, 59, S857
%
\bibitem[Zhang et al.(2001)]{ZhangEtal2001}
Zhang, J., Kundu, M. R., \& White, S. M. 2001, Sol. Phys., 198, 347

\end{thebibliography}
\end{document}